\documentclass[aip,reprint, amsmath, apl, amssymb]{revtex4-1}

\usepackage{graphicx}

\begin{document}

% Use the \preprint command to place your local institutional report number
% on the title page in preprint mode.
% Multiple \preprint commands are allowed.
%\preprint{}

% repeat the \author .. \affiliation  etc. as needed
% \email, \thanks, \homepage, \altaffiliation all apply to the current author.
% Explanatory text should go in the []'s,
% actual e-mail address or url should go in the {}'s for \email and \homepage.
% Please use the appropriate macro for the type of information

% \affiliation command applies to all authors since the last \affiliation command.
% The \affiliation command should follow the other information.

\title{Nonlinear THz Conductivity Dynamics in CVD-Grown Graphene}

% repeat the \author .. \affiliation  etc. as needed
% \email, \thanks, \homepage, \altaffiliation all apply to the current
% author. Explanatory text should go in the []'s, actual e-mail
% address or url should go in the {}'s for \email and \homepage.
% Please use the appropriate macro foreach each type of information

% \affiliation command applies to all authors since the last
% \affiliation command. The \affiliation command should follow the
% other information
% \affiliation can be followed by \email, \homepage, \thanks as well.
\author{Harold Y. Hwang}
\author{Nathaniel C. Brandt}
\affiliation{Department of Chemistry, Massachusetts Institute of Technology, Cambridge, MA 02139 USA}
\author{Hootan Farhat}
\author{Allen L. Hsu}
\author{Jing Kong}
\affiliation{Research Laboratory of Electronics, Massachusetts Institute of Technology, Cambridge, MA 02139 USA}
\author{Keith A. Nelson}
\affiliation{Department of Chemistry, Massachusetts Institute of Technology, Cambridge, MA 02139 USA}
\email[]{kanelson@mit.edu}

\begin{abstract}
We report strong THz-induced transparency in CVD-grown graphene where 92\%-96\% of the peak-field is transmitted compared to 74\% at lower field strength. Time-resolved THz-pump/THz-probe studies reveal that the absorption recovers in 2-3 ps. The induced transparency is believed to arise from nonlinear pumping of carriers in graphene which suppresses the mobility and consequently the conductivity in a spectral region where the light-matter interaction is particularly strong.
\end{abstract}

% insert suggested PACS numbers in braces on next line

\maketitle %\maketitle must follow title, authors, abstract and \pacs
The discovery of single-layer graphene has generated intense fundamental scientific and applications-based interest over the past several years. The characteristic linear electronic dispersion of the material gives rise to massless Dirac fermions \cite{GeimNatMat2007}. As a result, graphene exhibits unique carrier transport properties that are of interest in electronics and optics applications. Advances in large-area, single-layer graphene fabrication have shown promise for the development of practical graphene devices \cite{LiScience2009}. Large area graphene fabrication also makes terahertz (THz) measurements simpler to perform compared with exfoliated graphene where the sample size is much smaller, since THz spot sizes are much larger than typical exfoliated graphene samples.

The dc to low-frequency (through THz frequencies) conductivity of graphene is large compared to the conductivity at higher frequencies (mid-IR to visible), and several theoretical investigations have predicted extraordinary effects from both electronic and electromagnetic stimuli, including nonlinear frequency conversion and nonlinear conductivity effects due to the strong interaction of low-frequency light with graphene  \cite{BaoPCM2009, BaoPhysLettA2010, MikhailovEPL2007, MikhailovJPCM2008, MikhailovMJ2009, WrightAPL2009}. Ultrafast studies of photoexcited carriers in graphene with visible to near-IR pumping and visible to THz probing have revealed various aspects of inter and intra-band conductivity \cite{DawlatyAPL2008, GeorgeNanoLett2008, ChoiAPL2009, Lee2010}. However, excitation in the visible to near-IR range is relatively inefficient since the absorption at these frequencies is weak \cite{MakPRL2008}. Consequently, carrier  temperatures achieved in these experiments were only 100-200 K above equilibrium \cite{ChoiAPL2009}, whereas excitation with intense THz pulses is predicted to yield much higher temperature excursions \cite{BaoPCM2009, BaoPhysLettA2010}. Furthermore, the creation of electron-hole pairs in graphene with optical excitation leads to complicated relaxation dynamics related to electron-hole recombination, making interpretation of these experiments difficult at the microscopic level.

Tabletop generation of THz pulses with microjoule energies and field amplitudes of hundreds of kV/cm \cite{YehAPL2007} has enabled nonlinear electronic spectroscopy of conventional semiconductors \cite{HoffmannJOSAB2009, HoffmannPRB2009, HeblingPRB2010}. Here we present nonlinear THz transmission experiments that demonstrate THz-induced transparency in graphene and we study the dynamics of the nonlinear response with THz-pump/THz-probe measurements.

\begin{figure}[h!]
\includegraphics[scale=0.28]{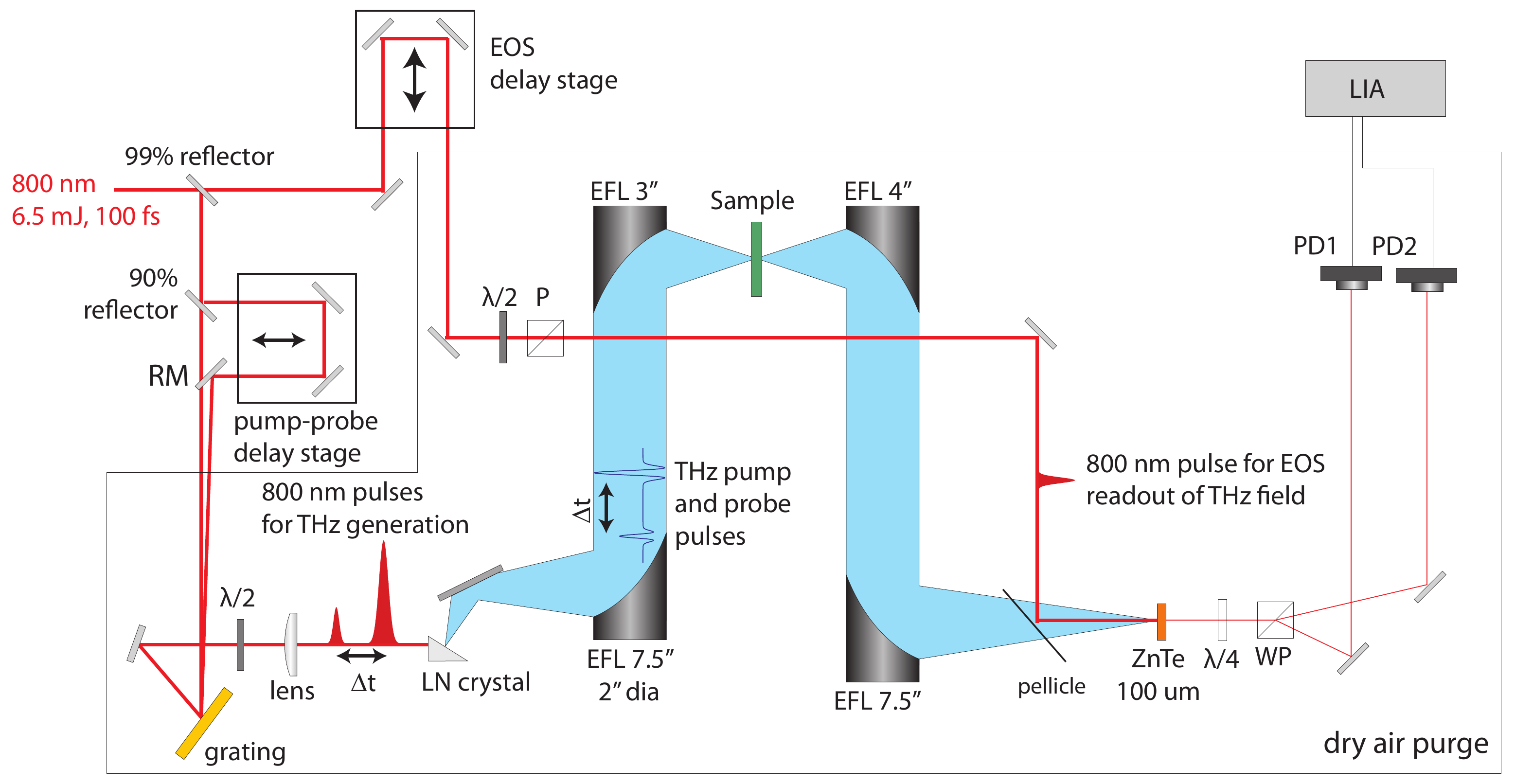}%
\caption{\label{}Experimental setup for both THz transmission and THz-pump THz-probe spectroscopy. In transmission experiments, the probe arm is blocked. The transmitted THz field is overlapped with a variably delayed 800-nm readout pulse in a ZnTe electro-optic sampling (EOS) crystal, and the THz-induced depolarization of the readout pulse reveals the time-dependent THz field profile. EOS, electro-optic sampling; RM recombination mirror; EFL, effective focal length; LN, lithium niobate; P, polarizer; WP, Wollaston prism; PD, photodiode;  $\lambda$/2, half waveplate;  $\lambda$/4, quarter waveplate; LIA, lock-in amplifier.}
\end{figure}

The experimental setup is shown in Figure 1. High-field THz pulses were generated by optical rectification with tilted pulse front (TPF) excitation in lithium niobate (LN), giving pulse energies in excess of 3 $\mu$J at 1 kHz repetition rate \cite{YehAPL2007, FeurerARMS2007, HeblingOpticsExpress2002}. THz pulses were collimated and focused onto the sample with a pair of off-axis parabolic mirrors, and the transmitted THz light was imaged onto a ZnTe electro-optic sampling crystal for detection. THz pulse intensities were varied with wiregrid polarizers for nonlinear transmission measurements. The peak electric field in the setup is estimated to be over 100 kV/cm at the sample.

THz-pump/THz-probe measurements were performed in a collinear geometry by splitting the optical pulse used for THz generation into pump and probe pulses with an adjustable time delay between them and then recombining them in a common LN crystal \cite{HoffmannJOSAB2009, HoffmannPRB2009, HeblingPRB2010}. \begin{figure}[h!]
\includegraphics[scale=.26]{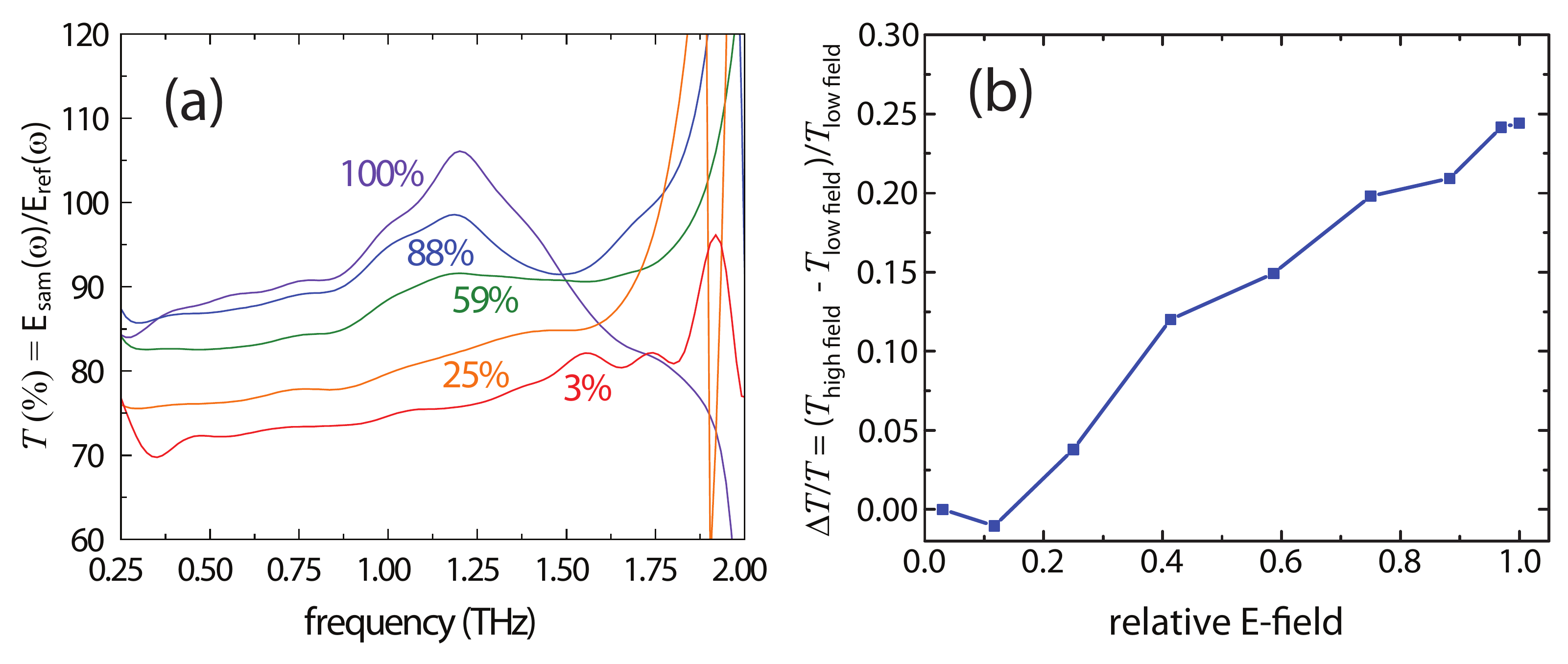}%
\caption{\label{}(a) THz field percent transmission at different percentages of the maximum THz field strength in our system. The transmission increases as THz field strength increases. (b) Peak field transmission versus relative E-field normalized to peak field transmission at the lowest THz field.}
\end{figure} The transmitted THz fields were measured by electro-optic sampling in ZnTe. For the time-resolved measurements, the peak amplitude in the transmitted THz probe field profile is reported as a function of THz pump-probe delay.

Graphene samples were grown on a copper substrate by CVD \cite{LiScience2009} and transferred to either fused silica or high resistivity silicon substrates.  The graphene covered roughly half of the substrate area, which allowed us to record reference and sample scans with a common substrate. Hall effect measurements on several device sizes indicate a sheet hole concentration of about 5 x $10^{12}$ cm$^{-2}$ for either fused silica or silicon substrate samples. Based on a linear density of states and a Fermi-Dirac thermal distribution\cite{FengAPL2007}, the Fermi energy was $E_{f}$ = -270 meV at room temperature. Hole-doped graphene at our carrier sheet densities should only exhibit intraband responses since the calculated Fermi energy far exceeds photon energies of our THz pulses. All THz spectroscopic measurements were taken at room temperature.

Figure 2a shows spectrally resolved THz field transmission ($T = E_{sam}(\omega)/E_{ref}(\omega)$) for various THz field strengths as a function of frequency ($\omega$), indicating a strong increase in transmission with increasing field. This can be attributed to a decrease in carrier mobility as THz excitation redistributes carrier energies within the conduction band \cite{BaoPCM2009, BaoPhysLettA2010}. In this case, the largest effect comes from heating the holes with the THz pulse since the sample is strongly hole doped. There is a notable feature that grows in at 1.2 THz as the THz field strength increases, exceeding 100\% transmission at the highest incident field strength which suggests gain in this frequency range.

Figure 2b shows THz peak field transmission normalized to peak field transmission at the lowest THz field strength ($ \Delta T/T = (T_{high field} - T_{low field})/T_{low field}$ where $T = max(E_{sam})/max(E_{ref})$ and $E_{sam}$ is the electric field from the graphene on the substrate and $E_{ref}$ is the electric field from the substrate alone) as a function of THz field strength. The relative increase to 25\% at the highest THz field strength indicates significant nonlinearity over the range of field strengths measured. The corresponding peak field transmission goes from 74\% to 92\% from low to high THz field strength (which is 54\% to 82\% when integrating the intensity of the measured THz pulse $\int E_{sam}^2 (t)dt/\int E_{ref}^2 (t)dt$).

\begin{figure}[h!]
\includegraphics[scale=0.30]{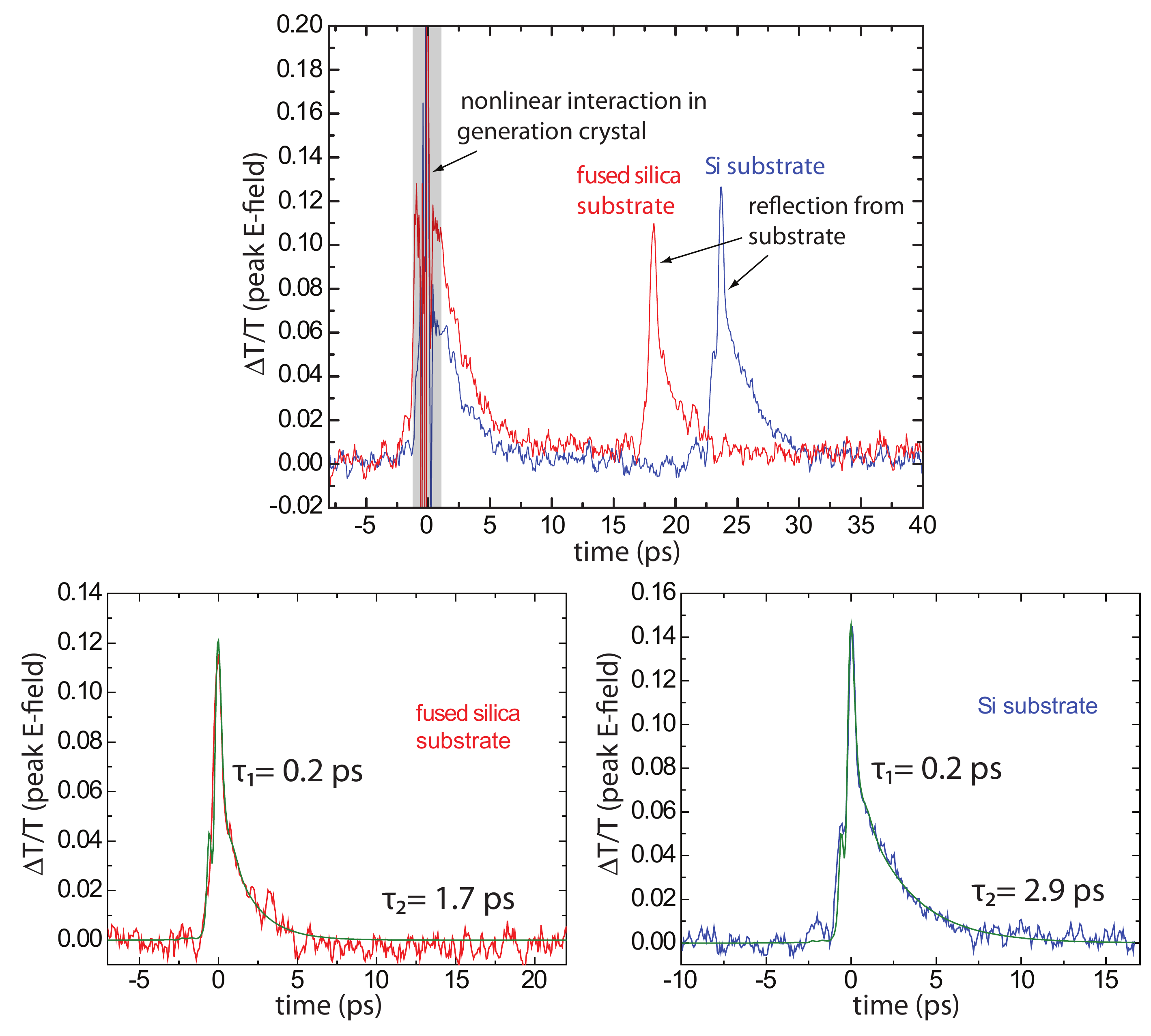}%
\caption{\label{}(a) THz-pump THz-probe peak field scans for graphene on fused silica (red) and silicon (blue). The EOS delay is set so that the optical readout pulse is temporally overlapped with the peak of the transmitted THz probe pulse, and the THz pump-probe time interval is varied. The large signals at time zero arise from the nonlinear interaction of the optical pump pulses in the THz generation crystal. The second signals are due to reflection of the THz pump pulse in the sample substrates; the time delays are consistent with the THz refractive indices in fused silica and silicon. (b) and (c) The signal after the first reflection of the pump in the substrate of each sample is shown, with the zero of time reset to match the overlap between the probe pulse and the internally reflected pump pulses. This allows us to examine signals with time-coincident THz pump (after reflection) and probe pulses that were not generated at the same time in the LN crystal, avoiding nonlinear optical effects on the sample. The data were fit to a convolution of the square of the THz pump field with a biexponential decay.}
\end{figure}

In time-resolved THz pump-probe measurements, the signal at t = 0 was not reliable because this corresponded to the two optical pulses that generated the THz pump and probe pulses overlapping inside the LN crystal where they could interact nonlinearly, influencing the THz generation process. However, there was a significant back-reflection of the THz pump pulse from the substrate-air interface (~35\% and 55\% reflection of the THz field per interface in fused silica and silicon, respectively), and the reflected THz pump pulse was sufficiently intense to produce a significant change in probe pulse transmission. Our signal from each sample therefore consists of two temporally separated components: a signal that starts at t = 0, which is only reliable at time delays greater than the THz pulse duration of about 1 ps, and another signal that starts when the THz pump-probe delay time matches the THz round-trip time $t_{RT}$ (roughly 20 ps) inside the sample. Since the second signal component permits reliable measurement of the sample response at short probe delay times relative to $t_{RT}$, we use this component for our analysis. In principle, the signal would depend on sample responses from the first pass of the THz pump pulse through the graphene layer at t = 0 as well as the second pass at $t = t_{RT}$, but the sample response does not appear to persist for nearly that long.

Time-resolved THz-pump/THz-probe data shown in Fig. 3 reveal fast decay dynamics that exceed the time resolution given by the THz pulse, and a slower component that can be fit to an exponential decay with  $\tau$ = 1.7 ps for graphene on fused silica and $\tau$  = 2.9 ps for graphene on silicon.  The peak change in $\Delta T/T$ reaches 12-14\% (Fig. 3b-c) for both substrates. Absolute transmission of the THz probe pulse reaches 92\% and 96\% in the fused silica and silicon substrates, respectively, compared to ~85\% where the peak field transmission prior to the arrival of the pump pulse is. Absolute probe peak-field transmission may exceed 100\% at the true time zero suggesting that nonlinear THz frequency conversion may be possible in graphene.
The strong induced transparency that we measure is in contrast with expectations from a Drude-like treatment of the frequency-dependent intraband conductivity, $\sigma$, in graphene \cite{DawlatyAPL2008, GeorgeNanoLett2008, ChoiAPL2009, Lee2010}.

\begin{equation}
\dfrac{\sigma_{intra}(\omega)}{\sigma_Q} =\dfrac{8k_B \textit{\textbf{T}}}{\pi\hbar}ln\left(e^{\frac{-E_F}{2k_B \textit{\textbf{T}}}}+e^{\frac{E_F}{2k_B \textit{\textbf{T}}}}\right) \dfrac{1}{\omega^2 \tau+1/\tau}
\end{equation}

where $\sigma_{Q}$ is the universal dc quantum conductivity, $k_B$ is Boltzmann's constant, $\textit{\textbf{T}}$ is temperature, $\hbar$ is the reduced Planck's constant, $E_F$ is the Fermi energy, $\omega$ is frequency, and $\tau$ is the momentum scattering time ($\tau$ = 2 fs typically in graphene \cite{ChoiAPL2009, Lee2010}). Here the electron and hole responses are symmetric, such that electron- or hole-doped graphene should exhibit the same conductivity. The conductivity can be related to the transmission ($T$) by:

\begin{equation}
T=  \dfrac{1}{\left|1+\frac{Z_0 \sigma(\omega)}{n_s+1}\right|^{2}}
\end{equation}

Where $Z_0$ is the vacuum impedance and $n_s$ is the substrate refractive index. It is assumed that the THz pulse heats carriers which thermalize in a short time compared to the experimental time resolution ($\sim$100 fs). However, this model predicts induced absorption and not induced transparency at higher carrier temperatures.

Theoretical treatments of graphene under a dc bias or ac excitation at THz frequencies \cite{BaoPCM2009, BaoPhysLettA2010} have predicted a very strong decrease in carrier mobility and consequently a decrease in conductivity at even modest dc and THz field strengths ($\sim$1 kV/cm). These studies take into account damping due to electron-impurity and both electron-acoustic phonon and electron-optic phonon scattering at typical carrier densities (N = 0.5-1.5 x 10$^{12}$ cm$^{-2}$). By separating the mobility into impurity, acoustic phonon, and optic phonon contributions at various lattice temperatures, it is possible to assess which scattering mechanisms dominate in the conductivity. At low lattice temperatures, impurity scattering dominates. At moderate temperatures (T $\sim$ 300 K), acoustic phonon scattering becomes significant, and finally at still higher temperatures (T $>$ 700 K) optic phonon scattering becomes significant. For the nonlinear conductivity response, similar arguments explain the decrease in conductivity with increasing field strength since the relative change in electron temperature versus lattice temperature decreases with increasing lattice temperature and increasing field strength. As the acoustic and optical phonon modes are more populated at higher lattice temperatures, energy dissipation occurs more efficiently, limiting the electron drift velocity $v_d$ and thus the conductivity since $v_d = \mu E$ and $\sigma = Ne\mu$  (at high hole concentration assuming electron and hole mobilities are roughly equal) where $\mu$  is mobility, $E$ is applied electric field, $N$ is the number density of carriers, and $e$ is the charge of an electron.

We observed strong THz-induced transparency in CVD-grown graphene. We believe the effect is due to heating of holes by the THz pulse, which suppresses the carrier mobility through electron-phonon scattering processes. Time-resolved THz-pump/THz-probe spectroscopy indicated carrier cooling on a picosecond timescale, which is consistent with similar studies done with optical pump pulses \cite{DawlatyAPL2008, GeorgeNanoLett2008, ChoiAPL2009, Lee2010}. Further study is needed to fully explain the phenomena encountered in our experiments including a better theoretical treatment of scattering processes in graphene to fit our experimental conditions, and temperature dependent studies to investigate the effect of lattice temperature on such scattering processes.

The authors would like to acknowledge funding from ONR grants N00014-06-1-0459 and N00014-09-1-1103 and Dr. Keshav Dani and Prof. Mildred Dresselhaus for helpful discussions.
%\end{acknowledgments}

% Create the reference section using BibTeX:
\bibliography{graphene}

% Body of paper goes here. Use proper sectioning commands.
% References should be done using the \cite, \ref, and \label commands
%\section{}
%\label{}
%\subsection{}
%\subsubsection{}

% If in two-column mode, this environment will change to single-column format so that long equations can be displayed.
% Use only when necessary.
%\begin{widetext}
%$$\mbox{put long equation here}$$
%\end{widetext}

% Figures should be put into the text as floats.
% Use the graphics or graphicx packages (distributed with LaTeX2e).
% See the LaTeX Graphics Companion by Michel Goosens, Sebastian Rahtz, and Frank Mittelbach for examples.
%
% Here is an example of the general form of a figure:
% Fill in the caption in the braces of the \caption{} command.
% Put the label that you will use with \ref{} command in the braces of the \label{} command.
%
% \begin{figure}
% \includegraphics{}%
% \caption{\label{}}%
% \end{figure}

% Tables may be be put in the text as floats.
% Here is an example of the general form of a table:
% Fill in the caption in the braces of the \caption{} command. Put the label
% that you will use with \ref{} command in the braces of the \label{} command.
% Insert the column specifiers (l, r, c, d, etc.) in the empty braces of the
% \begin{tabular}{} command.
%
% \begin{table}
% \caption{\label{} }
% \begin{tabular}{}
% \end{tabular}
% \end{table}

% If you have acknowledgments, this puts in the proper section head.
%\begin{acknowledgments}
% Put your acknowledgments here.
%\end{acknowledgments}

% Create the reference section using BibTeX:
%\bibliography{your-bib-file}

\end{document}